\documentclass[12pt]{article}
\usepackage{epsf}
\usepackage{cite}
\usepackage{amsmath,amssymb}
\usepackage{graphicx}
\usepackage{color}
\usepackage{comment}

\usepackage{bm}

\begin{document}

\newcommand{\lsim}{\stackrel{<}{_\sim}}
\newcommand{\gsim}{\stackrel{>}{_\sim}}

\newcommand{\rem}[1]{{$\spadesuit$\bf #1$\spadesuit$}}
\newcommand{\TT}[1]{\textcolor{blue}{#1}}

\renewcommand{\theequation}{\thesection.\arabic{equation}}
\renewcommand{\thefootnote}{\fnsymbol{footnote}}
\setcounter{footnote}{0}
\def\thefootnote{\fnsymbol{footnote}}

\begin{titlepage}

\begin{center}

\vskip .75in

{\Large \bf   Probing Small Scale Primordial Power Spectrum \\ with  21cm Line Global Signal}
\vskip .75in

{\large
Shintaro Yoshiura$\,^{a}$, Keitaro~Takahashi$\,^{b,c}$, and Tomo~Takahashi$\,^{d}$
}

\vskip 0.25in

{\em
$^{a}$The University of Melbourne, School of Physics, Parkville, VIC 3010, Australia
\\
$^{b}$Faculty of Advanced Science and Technology, Kumamoto University, Kumamoto, Japan
\\
$^{c}$International Research Organization for Advanced Science and Technology, Kumamoto University, Japan
\\
$^{d}$Department of Physics, Saga University, Saga 840-8502, Japan
}

\end{center}
\vskip .5in

\begin{abstract}

We argue that the global signal of neutral hydrogen 21cm line can be a powerful probe of primordial power spectrum on small scales. Since the amplitude of small scale primordial fluctuations is important to determine the early structure formation and the timing when the sources of Lyman $\alpha$ photons are produced, they in turn affect the neutral hydrogen 21cm line signal. We show that the information of the position of the absorption trough can severely constrain the small scale amplitude of primordial fluctuations once astrophysical parameters relevant to the 21cm line signal are fixed. We also discuss how the uncertainties of astrophysical parameters affect the constraints.  

\end{abstract}

\end{titlepage}

\renewcommand{\thepage}{\arabic{page}}
\setcounter{page}{1}
\renewcommand{\thefootnote}{\#\arabic{footnote}}
\setcounter{footnote}{0}

\section{Introduction} \label{sec:intro}

Primordial density fluctuations give us essential information on the very early epoch of the Universe,
especially, on the inflationary era since they are considered to originate from fluctuations generated during inflation.
The nature of primordial fluctuations are now well measured by cosmological observations such as cosmic microwave background (CMB) from Planck \cite{Aghanim:2018eyx,Akrami:2018odb}. 
However, the scales observed by CMB correspond to the modes which exited the horizon  at around $N \sim 50-60$, with $N$ being the
number of $e$-folds counted  backward from the end of inflation, and only large scales with the range of several $N$  can be probed. 
Therefore to understand the dynamics of  inflation over a broad range, we also need to look into smaller scales than those observed by CMB.

It has been discussed that primordial black hole \cite{Bugaev:2008gw,Josan:2009qn,Emami:2017fiy,Sato-Polito:2019hws}, ultracompact minihalos \cite{Bringmann:2011ut,Emami:2017fiy}, CMB $\mu$ distortion \cite{Chluba:2012we},
and so on can probe primordial density fluctuations on much smaller scales 
although their constraints on the amplitude are rather weak compared to those obtained by CMB.
In this paper, we argue that small scale primordial fluctuations can also be probed by the global signal of neutral hydrogen  21cm line. 

Since the amplitude of primordial fluctuations changes how structure formation proceeds, which affects the timing when Lyman $\alpha$ photon sources are switched on,
then in turn, changes the structure of the global signal of neutral hydrogen 21cm line. 
Indeed, in our previous work \cite{Yoshiura:2018zts},  we have derived constraints on the so-called running parameters, which are commonly used to characterize higher order 
scale-dependence of primordial power spectrum and denoted as $\alpha_s$ and $\beta_s$, 
using this argument in the light of recent report from the EDGES low-band result \cite{Bowman:2018yin} where the absorption trough is detected at the frequency of $\nu = 78~{\rm MHz}$, corresponding to 
the redshift of $z \sim 17.2$\footnote{
 See, e.g.  \cite{Hills:2018vyr,2018ApJ...858L..10D,Bradley:2018eev,Singh:2019gsv,Spinelli:2019oqm}, for discussion on the analysis and interpretation of the EDGES result.
}.  In  \cite{Yoshiura:2018zts}, we have shown that, by just using the information of the position (frequency) of the absorption trough, one can derive the constraint on $\alpha_s$ and $\beta_s$ 
whose uncertainties are comparable to Planck\footnote{
For the analysis on expected constraints on the runnings from future 21cm fluctuations, see \cite{Kohri:2013mxa,Munoz:2016owz,Sekiguchi:2017cdy}.
}. Here it should be noted that, although it is very common to parametrize primordial power spectrum with the spectral index $n_s$ and its runnings $\alpha_s, \beta_s$, 
these parameters are usually defined at a reference wave number on large scale, which means that we extrapolate the power spectrum to smaller scales. 
This description would be valid when the whole observable scales are described by a smooth inflaton potential, however, for example, if multi-fields are responsible for primordial fluctuations and large and small scale fluctuations are given by different potential/sources, or if there are some local features in the inflaton potential and its power spectrum gets modified due to the local structure, small scale fluctuations cannot be described by the extrapolation from large scale ones. Therefore, if one considers a broad class of models, it would be preferable to probe small scale fluctuations without resorting to the extrapolation from large scale power spectrum.

In this spirit, we in this paper investigate to what extent we can directly probe the amplitude of primordial power spectrum (PPS) on small scale by using the global signal (GS) of 21cm line.
For this purpose, we directly constrain the amplitude of small scale primordial fluctuations  from the feature of the absorption trough of the global signal of 21cm line, particularly having the result from EDGES in mind. We utilize not only low-band data \cite{Bowman:2018yin}, but also high-band one \cite{Monsalve:2017mli} from which we can obtain constraints on the amplitude of small scale fluctuations, but also clearly see what scales (wave number) mainly affect the 21cm signal. Such an analysis would also bring us great insight towards future observations of 21cm line. {We note that the result of EDGES low-band  still attracts a lot of debate, and therefore we just use the fact that the absorption trough is not detected except the redshift range of $14 < z < 22$ and set aside the depth of the absorption line reported by EDGES. We emphasize that primary purpose of this work is to illustrate the power of 21cm global signal to  constrain cosmology such as the primordial power spectrum. }

The structure of this paper is as follows. In the next section, we briefly summarize how we calculate the 21cm global signal relevant to this work and 
how we model primordial power spectrum to characterize small scale fluctuations. 
Then in Section \ref{sec:result}, we present our results on constraints on the amplitude of small scale primordial power spectrum. 
Some details of our analysis and uncertainties due to astrophysics are also discussed there.
The final section is devoted to summary of this paper. We assume a flat $\Lambda$CDM cosmology with ($\Omega_m$,$\Omega_\Lambda$,$\Omega_b$,$h$,$n_s$,$\sigma_8$)=(0.31, 0.69, 0.048, 0.68,0.97,0.81) \cite{Ade:2015xua} where 
$\Omega_m, \Omega_\Lambda$ and $\Omega_b$ are density parameters for matter, a cosmological constant $\Lambda$ and baryon, $h$ is the Hubble parameter normalized by $100~{\rm km}/{\rm sec}/{\rm Mpc}$, 
$n_s$ is the spectral index and $\sigma_8$ is the amplitude of matter fluctuations in 8~Mpc$/h$ sphere.

\section{21cm global signal and modeling of primordial power spectrum} \label{sec:model}

\subsection{\label{sec:level2} 21cm global signal and 21cmFAST}

The observable of the neutral hydrogen 21cm line is the 21cm brightness temperature $\delta T_b$, which is defined as an emission or an absorption against to the {radiation background}. The $T_b$ can be written by (see, e.g. \cite{Furlanetto:2006jb})
\begin{equation}
\label{eq:T_b}
\delta T_b(z) \simeq 27\, x_{\rm HI} \left( \frac{\Omega_bh^2}{0.023} \right) \left( \frac{0.15}{\Omega_m h^2} \right)^{1/2}  \left( \frac{1+z}{10} \right)^{1/2} 
\left( \frac{T_S - T_{\rm rad}}{T_S} \right)\, {\rm mK}\,  , 
\end{equation}
where $x_{\rm HI}$ is the neutral fraction of hydrogen, $T_S$ is the spin temperature, $T_{\rm rad}$ is the temperature of radiation background. {We note that although CMB is typically regarded as the radiation background, but strong radio background from high-$z$ objects is also suggested to explain the EDGES result \cite{2018ApJ...858L..17F,2018PhRvL.121a1101F,2019MNRAS.tmp.3136E,2019arXiv191010171M}.} The spin temperature evolves through the coupling to the kinetic temperature of intergalactic medium (IGM) via Lyman-$\alpha$ photons. The 21cm signal is observed as an absorption when $T_S$ is lower than $T_{\rm rad}$, and becomes an emission once the gas is heated sufficiently by the X-ray photons and $T_s$ gets larger than $T_{\rm rad}$.

For the analysis in this paper, we use the latest version of 21cmFAST \cite{Park:2018ljd, Mesinger:2010ne} for calculating the 21cm line global signal, which we briefly describe here. The 21cmFAST solves the reionization, the evolution of the spin temperature and evaluate the 21cm brightness temperature distribution. The matter density distribution is generated by using Zel'Dovich approximation from a high resolution initial condition. The ionization is computed by comparing the number of the ionizing photon and the number of hydrogen atom within a sphere. To compute the evolution of spin temperature, the Lyman-$\alpha$ and X-ray background are calculated using approximations such as a step function for the optical depth and the use of the excursion set theory.  The simulation is performed with a volume of $\rm (160Mpc)^3 $ and $128^3$ grids. The heating and ionization are solved from $z=50$.  For further details,  see \cite{Park:2018ljd, Mesinger:2010ne}.

Various astrophysical parameters are employed in the 21cmFAST: the fraction of baryon gas in stars is given by a power-form as a function of halo mass $M_h$ as  $f_\ast (M_h) = f_{\ast,10} (M_h/10^{10} M_\odot)^{\alpha_\ast}$ where $f_{\ast,10}$ is the value at $M_h=10^{10}M_\odot$ and $\alpha_\ast$ is its power-law index. The escape fraction of ionizing photon is parameterized in the same way as $f_{\rm esc}(M_h) = f_{\rm esc,10} (M_h/10^{10} M_\odot)^{\alpha_{\rm esc}}$ with $f_{\rm esc,10}$ and $\alpha_{\rm esc}$ being the value at $M_h=10^{10}M_\odot$ and the power-law index. There are also parameters describing the minimum energy of X-ray photon $E_0$ and X-ray luminosity per star formation rate $L_X$. Star formation rate per stellar mass is characterized by the time-scale $t_*$ in units of the Hubble time, which is treated as a free parameter. In small galaxies, star formation might be suppressed by photo feedback and less gas accretion. To handle these effects, the star formation in halo is reduced by $\exp({-M_{\rm turn}/M_h})$ where $M_{\rm turn}$ is the mass threshold under which halos cannot host a star-forming galaxy.

The absorption trough of the global 21cm signal strongly depends on the parameters related to X-ray emission and Lyman-$\alpha$ photons. In the 21cmFAST, high mass X-ray binaries is assumed as X-ray sources, and the fiducial value of $L_X$ is chosen based on composite spectral energy distribution (SED) model which is derived from population synthesis calculation in \cite{Fragos:2012vf, Das:2017fys}. In particular, the X-ray luminosity per star formation rate and the power law index of X-ray would significantly change the position of absorption trough, which are degenerate with the effect caused by the small scale amplitude of primordial power spectrum. However, these parameters would be strictly constrained by luminosity function and 21cm power spectrum \cite{Park:2018ljd}, which can remove the degeneracy between the small scale amplitude and astrophysical parameters.

Another important astrophysical parameter in deriving a constraint on the small scale amplitude of PPS  is the minimum virial temperature $T_{\rm vir}^{\rm min}$ which controls the minimum mass of halo $M_{\rm turn}$. In the latest version of the 21cmFAST, $M_{\rm turn}$ is used instead of $T_{\rm vir}^{\rm min}$. Since $M_{\rm turn}$ significantly affects the absorption trough in the global signal, this parameter would also be degenerate with that of the small scale amplitude of PPS. Therefore  we take several values of $M_{\rm turn}$ to show a constraint on the amplitude of PPS.  However, we note that $M_{\rm turn}$  can also be constrained by combining the information from 21 cm power spectrum and CMB \cite{Kern:2017ccn}, which could break the degeneracy.

The rest of other astrophysical parameters are fixed to the fiducial values of 21cmFASTv2: $f_*=0.05$, $\alpha_*=0.5$, $f_{\rm esc}=0.1$, $\alpha_{\rm esc}=-0.5$, $E_0=500~{\rm eV}$, $L_X=10^{40.5} \rm erg~s^{-1}~M_{\rm sun}^{-1}~yr$ and $t_*=0.5$. Cosmological parameters are fixed as given at the end of Section~\ref{sec:intro}. In principle, Markov chain Monte Carlo (MCMC) analysis where the astrophysical and cosmological parameters are varied could be possible by using, e.g., the 21CMMC \cite{Greig:2015qca,Greig:2017jdj}, however varying the astrophysical and cosmological parameters is computationally very demanding and could easily slower the calculation due to iterative generation of matter density distribution. But some astrophysical parameters such as $L_X$ would be degenerate with the small scale amplitude of PPS, we also discuss in some detail how other astrophysical parameters affect the constraints on the amplitude of small scales in Section~\ref{sec:result}. 

{While the 21cm line global signal can be calculated by using other codes such as ARES\cite{2014MNRAS.443.1211M} and such 1D codes might allow us to perform MCMC analysis, the 21cmFAST can calculate the 21cm power spectrum, as well as the global signal, which would also be a powerful tool to constrain the primordial power spectrum. In future work we plan to study the constraint by using both the 21cm global signal and the power spectrum. }

\subsection{\label{sec:level3} Primordial power spectrum and mass function}

The number of ionizing photons, the emissivities of X-ray photon and Lyman-$\alpha$ photon are evaluated using the halo mass function (HMF). Since the HMF depends on the amplitude of the PPS,  the 21cm GS is affected by the PPS through the HMF. In our analysis, we assume the following form for the PPS:
\begin{eqnarray}
P_{\rm prim} (k) &=& 
A_0 \left(\frac{k}{k_{\rm ref}} \right)^{n_{s}-1}p(k), 
\end{eqnarray}
where $A_0$ and $n_s$ are the amplitude and the spectral index at the reference scale $k_{\rm ref}$. Actually in our analysis, we normalize the amplitude to give $\sigma_8 = 0.81$ which automatically determines $A_0$ when we calculate the 21cm GS with the 21cmFAST.
$p(k)$ parametrizes the amplitude of PPS on small scales which is defined as 
\begin{eqnarray}
p(k)&=&
\begin{cases}
p_1 \quad (10< k <10^2~ \rm Mpc^{-1})\\
p_2 \quad (10^2< k <10^3~ \rm Mpc^{-1})\\
1~ \quad (\rm else) \,,
\label{eq:param1}
\end{cases}
\qquad \textrm{ [Parametrization I]}
\end{eqnarray}
in which $p_1$ and $p_2$ describe the amplitude of PPS for the scales of $k = 10 - 10^2~ {\rm Mpc}^{-1}$ and $10^2 - 10^3 ~{\rm Mpc}^{-1}$, respectively. In the following, we only change the amplitude over the scales of $10~{\rm Mpc}^{-1} < k < 10^3~{\rm Mpc}^{-1}$. On large scales where $k<{\cal O}(1)~{\rm Mpc}^{-1}$, the amplitude of PPS is relatively well constrained by CMB and large scale structure. On the other hand, the amplitude of very small scales with $k > 10^3 ~{\rm Mpc}^{-1}$ does not affect much the 21cm global signal, and hence we change the amplitude of PPS over the scales of $10~{\rm Mpc}^{-1} < k < 10^3~{\rm Mpc}^{-1}$.

From the viewpoint of inflationary models, it may be natural to consider the case where the PPS changes very smoothly and is almost unchanged over several orders of $k$, which motivates us to think of a different binning such that 
\begin{eqnarray}
p(k)&=&
\begin{cases}
p_3 \quad (10< k <10^3~ \rm Mpc^{-1})\\
1~ \quad (\rm else) \,.
\label{eq:param2}
\end{cases}
\qquad \textrm{ [Parametrization II]}
\end{eqnarray}
In the following analysis, we investigate  constraints on $p_1, p_2$ (for  Parametrization~I) and $p_3$ (for the Parametrization~II), 
particularly focusing on the upper bound on these parameters, namely, we mainly investigate the case where the small scale amplitude of PPS is enhanced. However, we should note that we can also get a lower bound from the argument of 21cm GS. Actually, since Parametrization~II assumes just one bin for $10 < k < 10^3~{\rm Mpc}^{-1}$, a constraint on $p_3$ becomes tighter compared to those on $p_1$ and $p_2$ and we can also obtain a lower bound on $p_3$ for some set of astrophysical parameters.

As mentioned above, we normalize the amplitude of PPS to give $\sigma_8=0.81$.
It should be noted here that even if we change $p(k)$, the normalization of PPS is unchanged since the scales of $k = 10 - 10^3~ {\rm Mpc}^{-1}$ is too small to affect the value of $\sigma_8$. We should also mention that  $\sigma_8$ calculated in 21cmFAST is stable within our parameter range for $p_1, p_2$ and $p_3$:  $0.1<p_1<30$,  $1.0<p_2<10^8$ and $0.1<p_3<30$\footnote{

As we will discuss in the next section, the 21cm GS is not so sensitive to the values of $p_2$ compared to other ones, we take a broad range for $p_2$. 
}. Therefore the $p_1, p_2$ and $p_3$ change relative amplitude only. 

The Press-Schechter HMF is employed in the 21cmFAST. Given the PPS,  the variance of the density perturbation is calculated as
\begin{eqnarray}
\sigma^2(R) = \int\frac{d^3{\bf{k}}}{(2\pi)^3} P(k) W^2(k,R),
\label{eq:sigma}
\end{eqnarray}
where $P(k)$ is the matter power spectrum at present time and $R$ is the scale in real space. 
As a filter function, the real space top-hat filter whose form in the Fourier space is given by $W(k,R)= 3(\sin(kR)-kR\cos(kR))/(kR)^3$ is adopted as the default setting of 21cmFAST.  However, the choice of filter is somewhat arbitrary as long as the filter is suitable to the cosmological model. Although the real space top-hat filter is commonly chosen, it is not clear whether the filter is suitable for the enhanced/suppressed PPS models we consider in this paper. Although the appropriate choice of filter has to be tested by comparing the HMF obtained from N-body simulation in various enhanced/suppressed PPS model, we consider some variations for filter functions to study constraints on the small scale amplitude of PPS.

In our analysis, we mainly adopt the so-called smooth-$k$ filter which has been proposed in \cite{Leo:2018odn} although we also make an analysis using the real space top-hat filter and the Gaussian filter which is given by $ W(k,R) = \exp (-k^2R^2/2)$.  The form of the smooth-$k$ filter is given by
\begin{equation}
\label{eq:smooth_k}
W(k,R)= \frac{1}{1+(kR)^{\beta}} \,,
\end{equation}
where the mass at the scale $R$ is $M(r)=\frac{4\pi}{3}\rho(cR)^3$, with $\rho$ being the average mass density, and $\beta$ and $c$ are free parameters of the filter. It has been shown in \cite{Leo:2018odn}  that the HMF calculated with this filter agrees well  with the N-body simulation even for models with suppressed matter power spectrum on small scales. Although the free parameters $\beta$ and $c$  should be fitted to the HMF obtained by N-body simulation,  it has been argued in \cite{Leo:2018odn} that these parameters can be set to match  the result obtained by other standard filter such as the top-hat one. Therefore we in this work assume the values of these parameters so that the predicted HMF approximately matches that obtained from the top-hat filter for the model with $p_1=p_2=1$ (or $p_3=1$) in redshift range of $0<z<30$, which leads us to choose them as $(\beta,c)=(4.8, 2.1)$. We note that the HMF with smooth-$k$ filter is slightly lower than that with top-hat filter at large mass scales. {We also note that Sheth-Tormen mass function is used in the 21cmFAST. However, we just note that some latest simulations show a different shape of mass function (e.g. \cite{2007MNRAS.374....2R,Trac_2015}). }

\begin{figure*}
\centering
\resizebox{160mm}{!}{
\includegraphics[width=10cm,bb= 0 0 864 576]{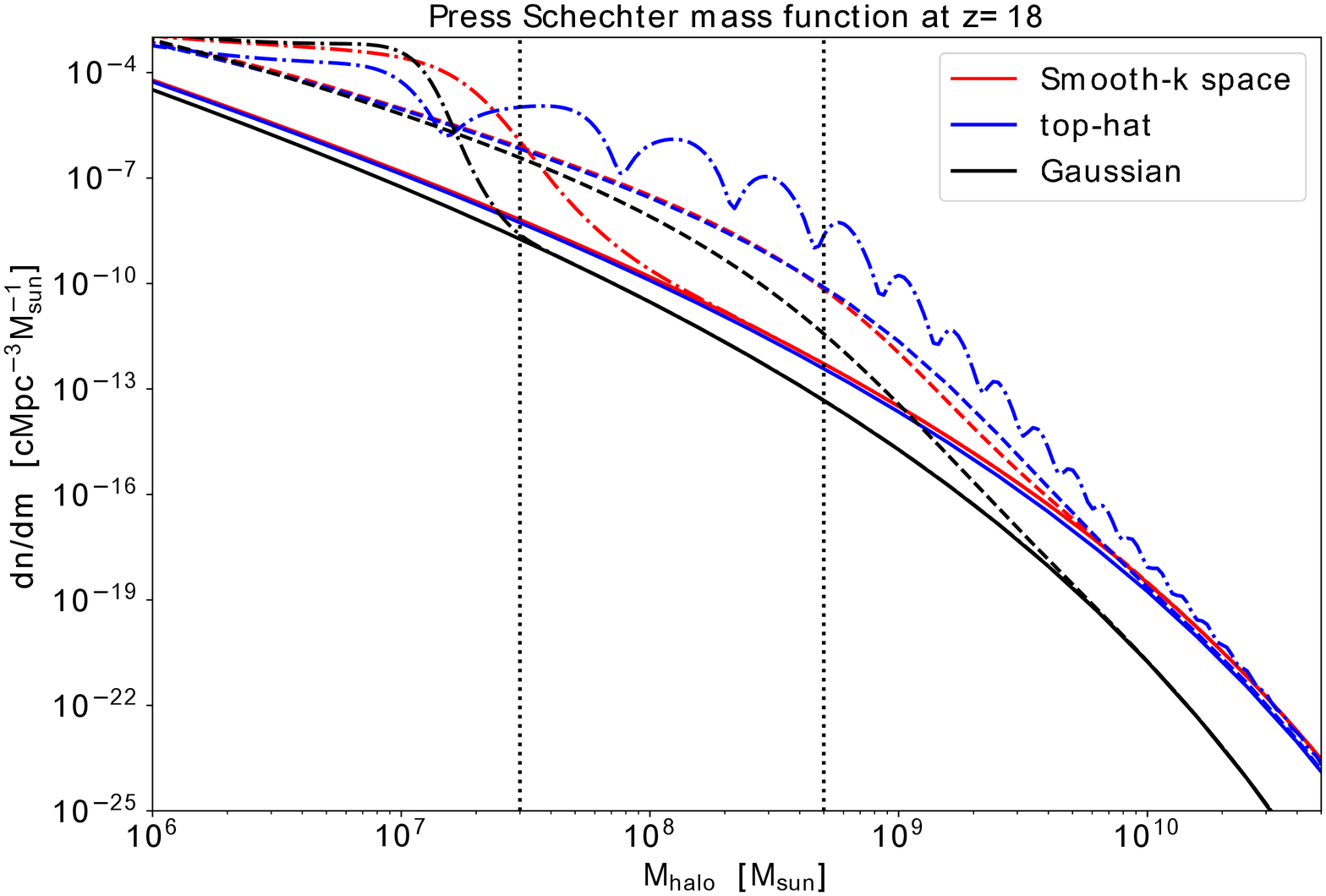} 
}
\caption{Press-Schechter halo mass function at $z=18$ using three types of filter function: the smooth-$k$ filter with $(\beta,c)=(4.8,2.1)$ (red), the top-hat filter (blue) and the Gaussian filter (black). Solid, dashed and dot-dashed lines are for the cases with $(p_1, p_2)= (1,1)$, $(p_1, p_2)= (5,1)$, and $ (p_1, p_2)= (1,10^5)$, respectively. The two vertical lines show the $M_{\rm turn} = 5\times 10^8 ~M_{\rm sun}$ and $3 \times 10^7~M_{\rm sun}$ which correspond to the fiducial and atomic cooling models, respectively, which are used in this work.  }
\label{fig:dndm}
\end{figure*}

Fig.~\ref{fig:dndm} shows the mass function at $z=18$ for these three filter functions.  Two vertical lines indicate mass threshold for the fiducial model with  $M_{\rm turn} = 5\times 10^{8} ~M_{\rm sun}$   and the atomic cooling one $ M_{\rm turn} = 3 \times 10^7 ~M_{\rm sun} $. Halos smaller than these threshold cannot contribute to radiation. We show the HMF for the cases with $p_1=p_2=1$ (solid lines), $p_1=5, p_2=1$ (dashed lines) and $p_1=1, p_2=10^5$ (dot-dashed lines). 
Since the Gaussian filter drastically reduces a contribution from small scales, the HMF is suppressed on all mass scales compared to the other filters, and becomes less sensitive to the values of $p_1, p_2$ and $p_3$. The top-hat filter gives an artificial feature for the case of a large value $p_2$. This oscillatory behavior is generated by the tail of the top-hat filter in $k$ space. To avoid this artificial behaviour, we adopt the smooth-$k$ filter as our fiducial filter.

\begin{figure*}
\centering
\resizebox{160mm}{!}{
\includegraphics[width=10cm,bb= 0 0 864 576]{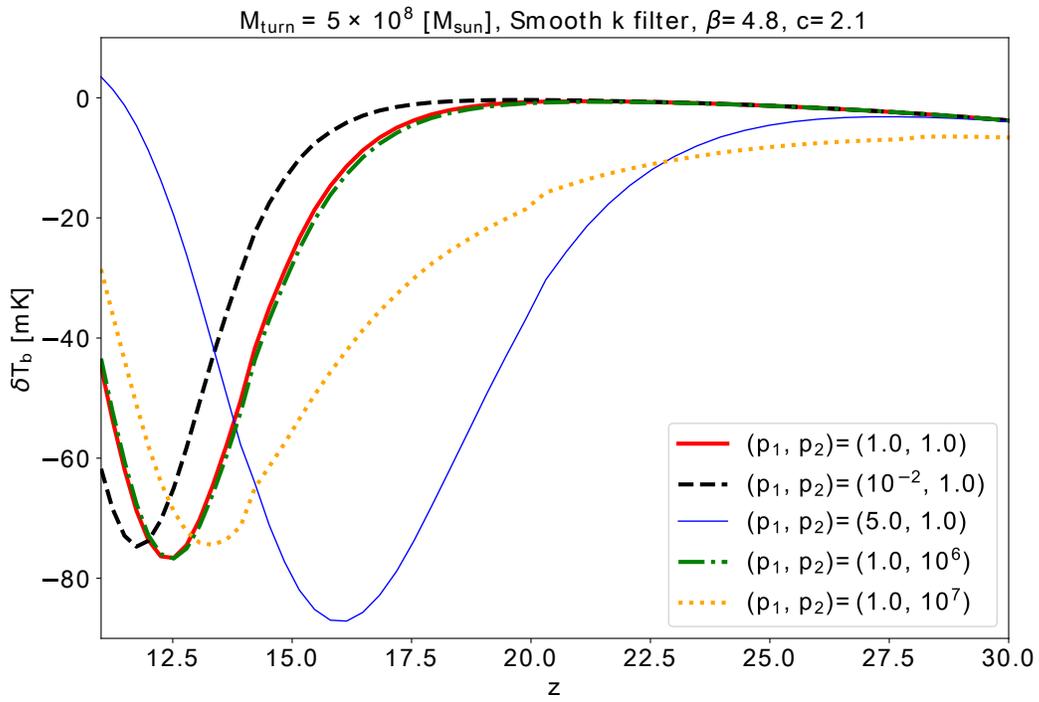} 
}
\caption{\label{fig:GS} Examples of 21cm global signal using the smooth-$k$ filter with $\beta=4.8$ and $c=2.1$ for several cases of $(p_1,p_2)$.}
\end{figure*}

Fig.~\ref{fig:GS} shows the evolution of $\delta T_b$ for five parameter sets of $p_1$ and $p_2$ in Parametrization~I. The minimum halo mass (mass threshold) is taken to be $ M_{\rm turn} = 5\times 10^8 ~M_{\rm sun}$ and the smooth-$k$ filter is adopted. Even for a small enhancement in $p_1$ such as the case of $(p_1,p_2)=(5,1)$, the absorption trough easily move toward higher redshift since the the number of high mass halos increases. On the other hand, the GS is not so sensitive to $p_2$ as seen from Fig.~\ref{fig:GS}.  The position (or redshift)  of the absorption trough moves only by $\Delta z<1$ even if we change $p_2$ from $1$ to $10^6$. This is because the enhanced mass scale of HMF is sufficiently smaller than the minimum halo mass in such a case.
Since $p_1$ can affect the number of halo on larger mass scales, the 21cm GS is sensitive to the value of $p_1$.  
It should be noted here that, the number of halo heavier than $M_{\rm turn}$  is essential for the 21cm GS. For example, for the smooth-$k$ filter with $(p_1, p_2)=(1,10^5)$, the HMF is enhanced only at $M_{\rm halo}\lesssim 10^8 ~ M_{\rm sun}$. In this case, the mass scale is much smaller than the threshold, and hence the halo cannot contribute to radiation. On the other hand, for the case of $(p_1, p_2)=(5,1)$  the mass function is enhanced at $M_{\rm halo}\sim 10^9 M_{\rm sun}$, which is larger than the mass threshold for fiducial model. Therefore,  large values of $p_1$ can easily affect the 21cm signal. However, we should note here that this kind of behavior depends on the filter function adopted. For example, if one adopts the top-hat filter and assuming $(p_1,p_2)=(1,10^5)$, the HMF is highly enhanced even at $M_{\rm halo}=10^9~ M_{\rm sun}$ which gives   the enhanced HMF and generate a large amount of photons, and hence X-ray heating happens at early redshift and the absorption trough moves toward higher redshift.

Recent EDGES result \cite{Bowman:2018yin} indicates  that the absorption trough is peaked at $z\approx 17$ with relatively flat spectrum for  $14<z<22$. Since, as already mentioned, the amplitude of PPS, especially on small scales, affects the GS of 21cm line, one can  constrain the amplitude by comparing the measured absorption spectrum and theoretical predictions with $p_1$ and $p_2$ or $p_3$ being varied. However, we should take account of the fact that the measured amplitude of 500~mK by EDGES cannot be realized without assuming some exotic physics such as baryon-dark matter interaction \cite{Barkana:2018lgd}. Therefore we only use  the information of the position of the absorption trough to constrain the small scale amplitude of PPS, and require that the absorption trough should appear at the redshift range of $14<z<22$. More specifically, when the global signal appears outside this redshift range with its depth deeper than 75~mK, which is  conservatively taken to be the threshold depth given the rms of residual is 25 mK in EDGES observation \cite{Bowman:2018yin}, we judge such a model is excluded.

In fact, models with the absorption trough peaked at $z<15$ can also be severely constrained by the EDGES high-band result \cite{Monsalve:2017mli} in which the rms of $\delta T_b$ is 17~mK at $6.5<z<15$ and 6~mK at $6.5<z<12$ after polynomial foreground removal. Thus, we also use the EDGES high-band result to derive the constraint in this paper. Since the EDGES high band data is affected by radio-frequency interference (RFI) and some systematics at $z>12$, we only adopt this constraint for models with absorption line at $z<12$. Since the systematic error is 35~mK \cite{Monsalve:2017mli}, we conservatively use 50~mK as the threshold value under which the model is considered to be excluded by the EDGES high-band result. It should be noted that the constraints are more affected by astrophysical parameters in this redshift ranges since the ionization parameters such as $f_{\rm esc}$ have a significant impact at such redshifts.

In table.~\ref{tab:tabthtabth}, we summary the thresholds and redshift ranges used in this work. When the 21cm global signal is lower than the threshold within the redshift range, we rule out such model.

\begin{table}[htb]
\begin{center}
\begin{tabular}{|r|l|l|}
  \hline
  Threshold & redshift & observation \\ 
  \hline \hline
  $-50$ mK & $z<12$  & Monsalve et al 2017 \cite{Monsalve:2017mli} \\ 
  \hline
  $-75$ mK & $13<z<14$  & Bowman et al 2018 \cite{Bowman:2018yin} \\
  \hline
  $-75$ mK & $22<z<30$  & Bowman et al 2018 \cite{Bowman:2018yin} \\ \hline
\end{tabular}
  \caption{Summary of the threshold and redshift ranges used for our constraints. Models are ruled out if the 21cm signal is lower than the threshold in the redshift range.}
    \label{tab:tabthtabth}
    \end{center}
\end{table}

\section{Constraints on the small scale amplitude  }\label{sec:result}

\subsection{Constraints on Parametrization I}\label{sec:param1}

In this section, we show  constraints on  $p_1$ and $p_2$ for the parametrization~I which represent the amplitudes at the scales of $10~{\rm Mpc}^{-1}<k<10^2~{\rm Mpc}^{-1}$ and $10^2~{\rm Mpc}^{-1} <k<10^3~{\rm Mpc}^{-1}$, respectively.

\begin{figure*}
\resizebox{160mm}{!}{
\includegraphics[width=10cm,bb= 0 0 648 576]{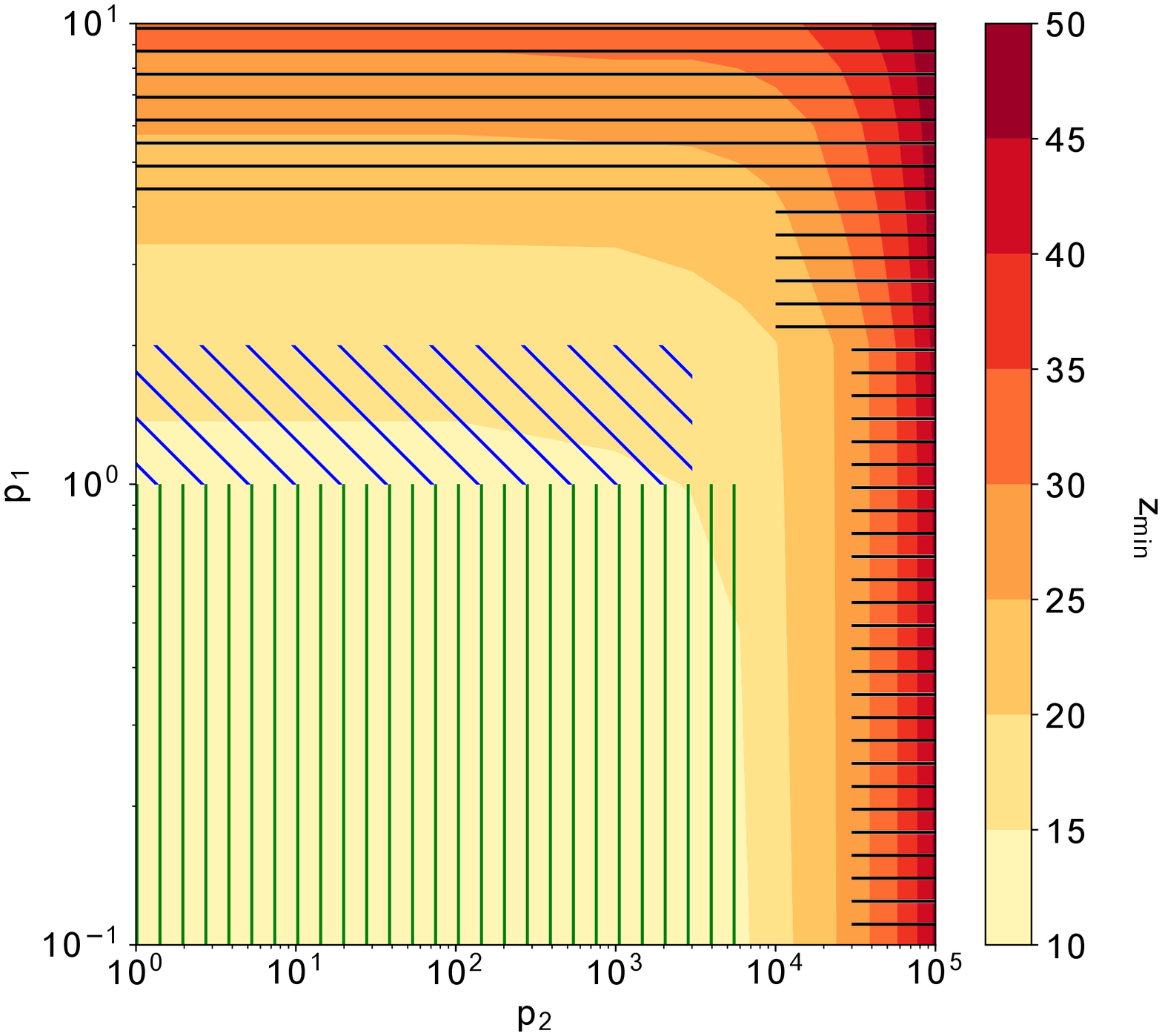} 
\includegraphics[width=10cm,bb= 0 0 648 576]{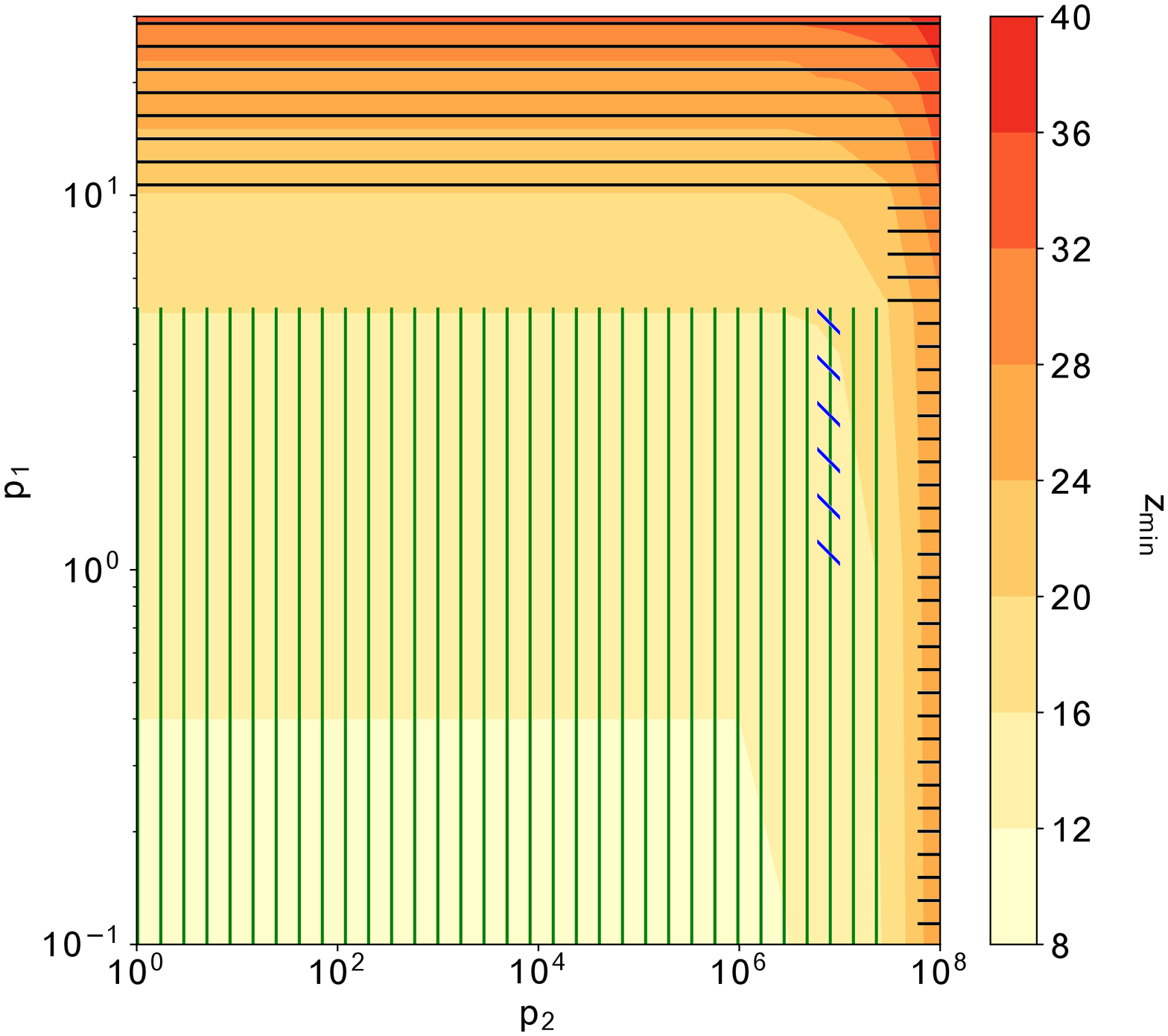}
}
\caption{\label{fig:result1} Constraints from the EDGES result and the central redshift of the absorption trough $z_{\rm min}$ are shown. We assume $M_{\rm turn} = 3\times 10^7 ~M_{\rm sun}$ (left panel) and $M_{\rm turn}=5\times 10^8~ M_{\rm sun}$ (right panel). Regions filled by horizontal line are excluded due to the deep absorption trough lies in $z>22$. Regions filled by diagonal line pattern are also ruled out due to the absorption line is at $z<14$. The regions filled by vertical line pattern indicate the models which are excluded by EDGES high band results. For large values of $p_1$ and $p_2$, the Lyman-$\alpha$ coupling and heating happen too early and hence such regions are excluded. On the other hand, lower values of $p_1$ and $p_2$ are not allowed due to the late X-ray heating. 
}
\end{figure*}

Fig.~\ref{fig:result1} depicts the constraints on $p_1$ and $p_2$. The regions filled by horizontal and diagonal lines are excluded due to the fact that the absorption trough lies at $22<z<30$ and $13<z<14$, respectively. As seen from the figure, $p_1$ can be strongly constrained since even a slight change in $p_1$ can affect the number of halo at large mass region as discussed in the previous section. Although the result depends on astrophysical parameters, the constraint on $p_1$ can be much tighter than other methods such as the one using primordial black holes around the scale of $k~\sim 10^{2}~{\rm Mpc}^{-1}$. For the case with $M_{\rm turn}=3\times 10^7 ~M_{\rm sun}$, the EDGES low-band results constrain the small scale amplitude of PPS as  $p_1 \lesssim 4$ and $p_2 \lesssim 3\times 10^4$. For the case with $M_{\rm turn}=5\times 10^8 ~M_{\rm sun}$, $p_1 \lesssim  10$ and $p_2 \lesssim  6\times 10^7$ are allowed.

In Fig.~\ref{fig:result1}, we also show the parameter range ruled out by the EDGES high-band data, which corresponds to the region filled by vertical line pattern. As mentioned in the previous section, we use the threshold of 50~mK and exclude a model if the amplitude of the absorption trough of 21cm GS is deeper than this value at $z<12$, which indicates that the heating and Lyman-$\alpha$ coupling should happen at early redshift. When the amplitude of PSS is suppressed, the structure formation is delayed and the heating and Lyman-$\alpha$ sources come later. This is the reason why the region with lower amplitude is excluded in Fig.~\ref{fig:result1}, which means that, interestingly,  the 21cm GS can provide lower limits on the amplitude of PPS.

It should be noted that in some models, such as those with $p_1\lesssim1$ and $p_2\lesssim1 \times 10^7$, the absorption trough is shallower than the threshold of 75~mK at all redshift and such models cannot be ruled out by EDGES low-band data. However, even in such cases, the EDGES high-band data can still provide a constraint on those models.

We also make a remark on the cases with large values of $p_1$ and $p_2$. As seen from Fig.~\ref{fig:result1}, when the values of $p_1$ and $p_2$ are large,  the position of the absorption trough lies at $z>30$  where there is no data from EDGES and one needs an instrument observing below 45~MHz to probe such redshifts. However such models can easily ionize the Universe and would be inconsistent with the optical depth observed by CMB. {Note that the EDGES low-band result  indicates the absorption line lies at $14 < z< 22$ and models with abosrption line at $z>30$ can be ruled out. However, we conservatively only use the non-detection feature to constrain the PPS and do not rule out such models. }

\begin{figure*}
\resizebox{160mm}{!}{
\includegraphics[width=10cm,bb= 0 0 648 576]{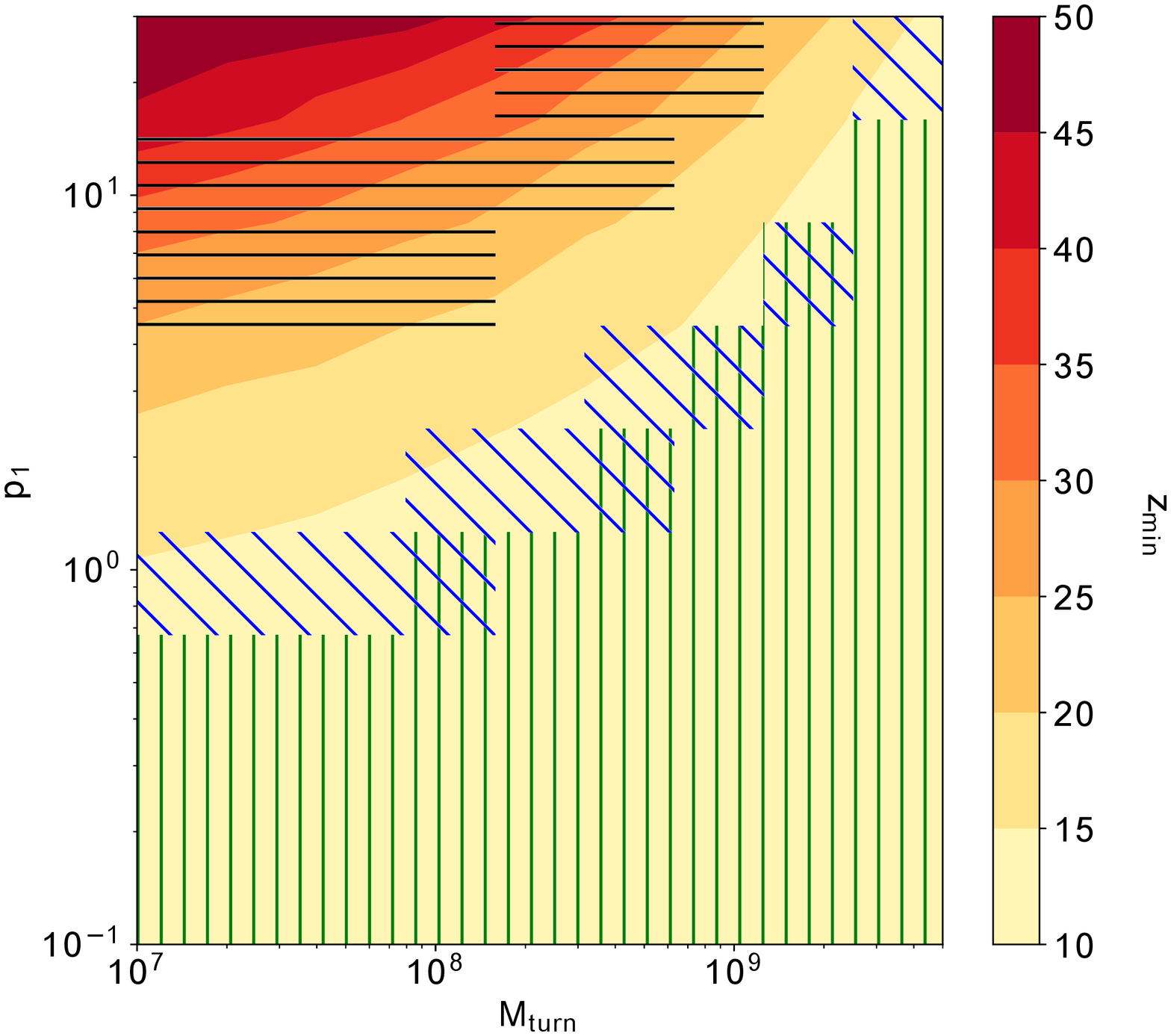} 
\includegraphics[width=10cm,bb= 0 0 648 576]{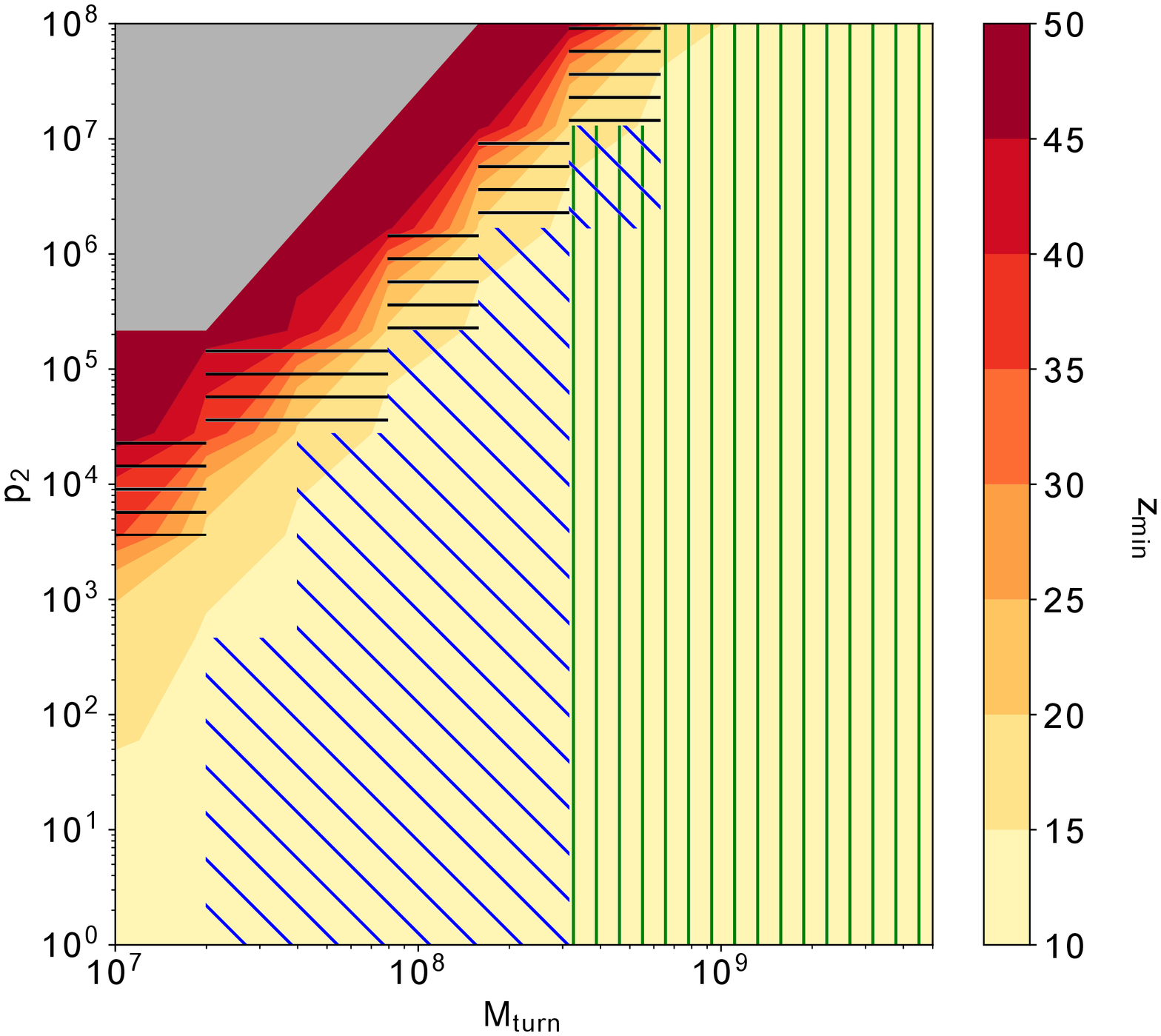}
}
\caption{\label{fig:result2} Allowed region and the redshift of the peak of absorption trough in  the $p_1$-$M_{\rm turn}$ (left) and the $p_2$-$M_{\rm turn}$ (right) planes.  We set $p_2=1$ and $p_1=1$ in the left and right panels, respectively. The region filled by vertical lines, horizontal lines and diagonal lines are same as Fig.~\ref{fig:result1}. In the figure, we can see correlation between $M_{\rm turn}$ and $p_1$ ($p_2$). At top left corner of the right panel, reionization has done before $z=50$ and the peak redshift cannot be defined. }
\end{figure*}

\begin{figure*}
\resizebox{160mm}{!}{
\includegraphics[width=10cm,bb= 0 0 648 576]{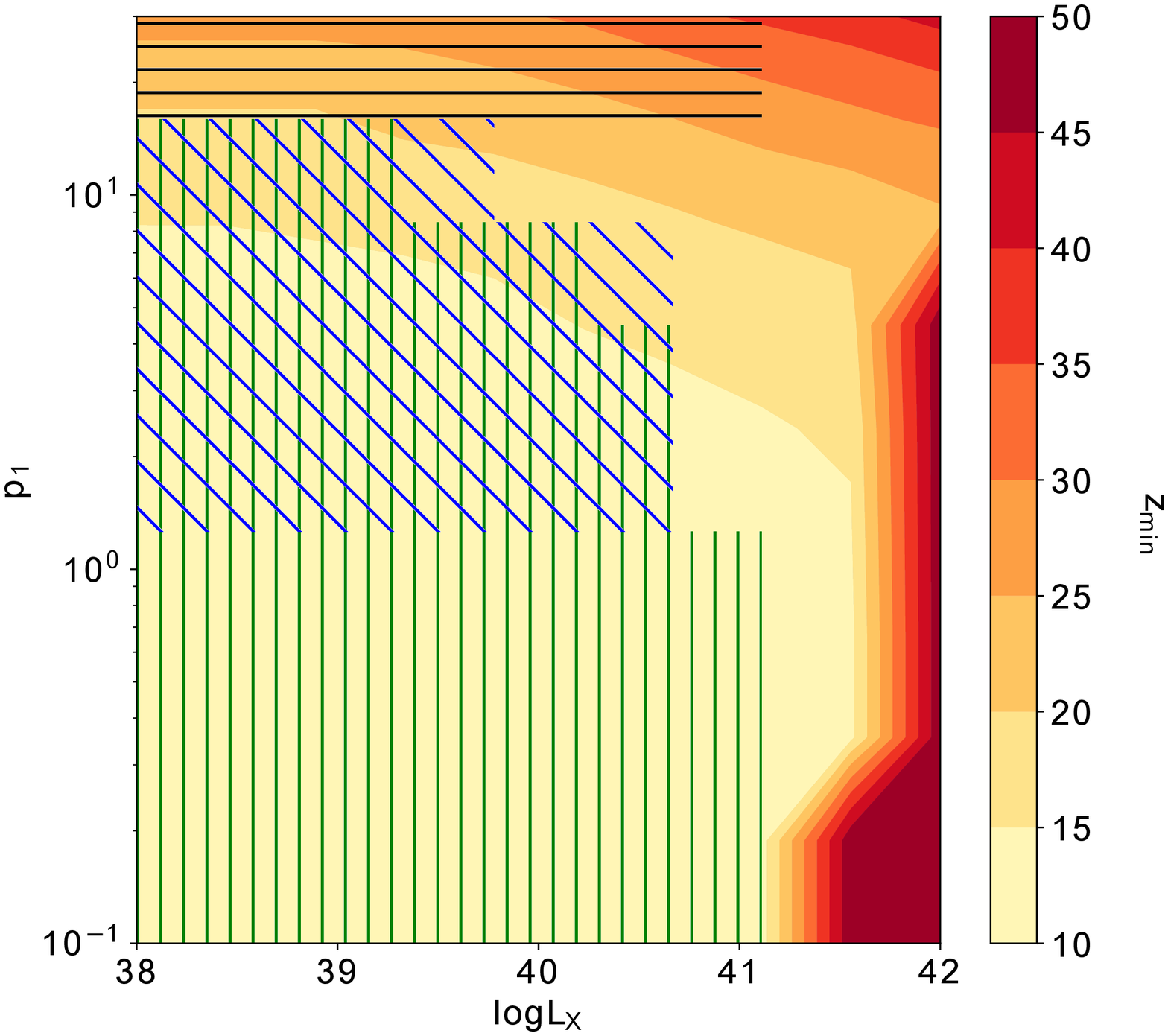}
\includegraphics[width=10cm,bb= 0 0 648 576]{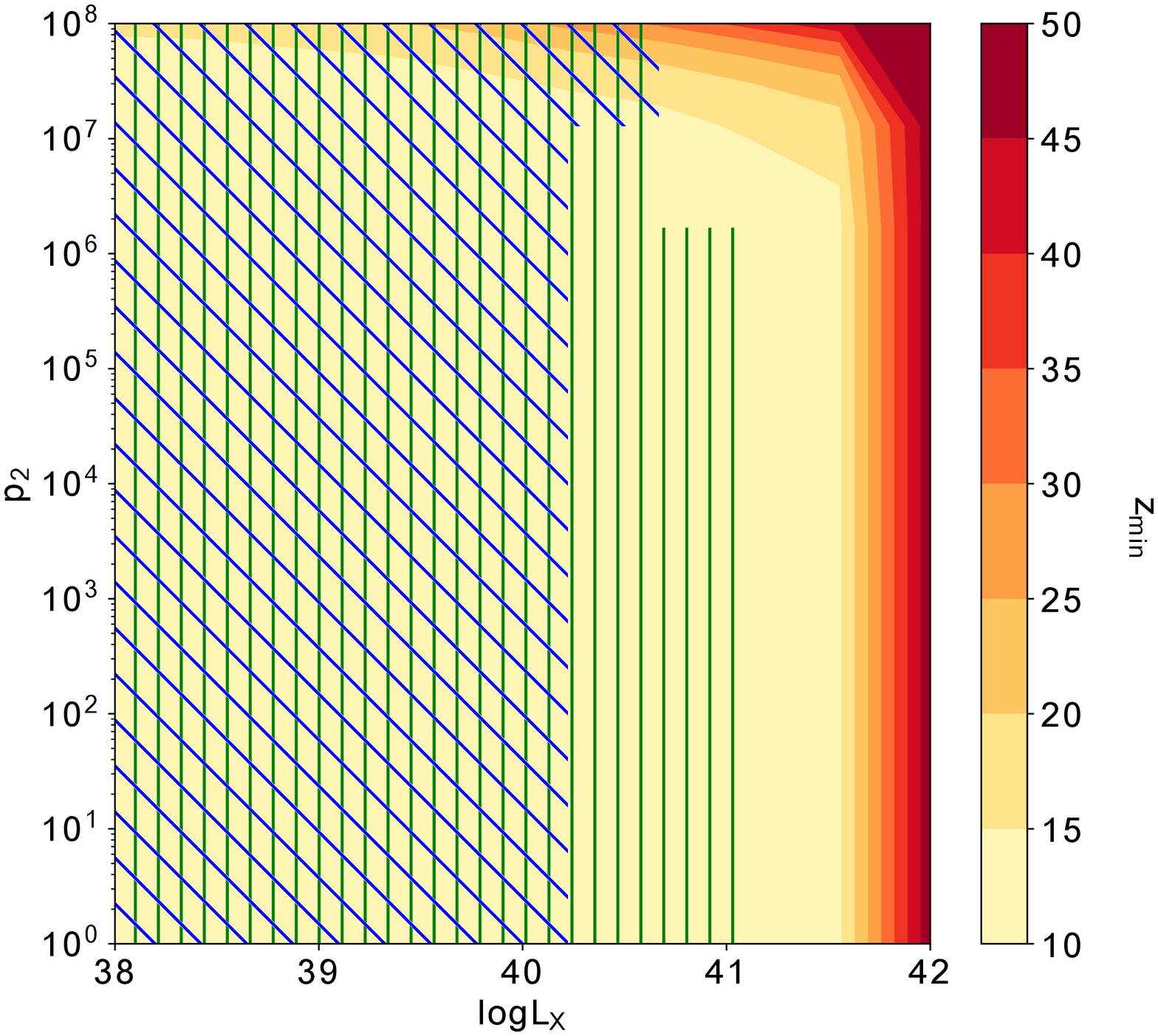}
}
\caption{\label{fig:p1_Lx} Constraints from EDGES on the $p_1$-$L_X$ (left) and $p_2$-$L_X$ (right) planes are shown in the same manner as Fig.~\ref{fig:result1}. 
}
\end{figure*}

Now we discuss how the assumption on astrophysical parameters affects the constraint on the small scale amplitude of PPS. As seen from Fig.~\ref{fig:result1}, the constraints on the $p_1$ and $p_2$ depend on $M_{\rm turn}$, and hence we can expect there exists some correlation between $(p_1, p_2)$ and $M_{\rm turn}$. Fig.~\ref{fig:result2} shows the constraint on $p_1~(p_2)$--$M_{\rm turn}$ plane with $p_2~(p_1)$  fixed to be unity. At the top left corner of the panels, Lyman-$\alpha$ coupling and X-ray heating happen very early since large values of $p_1 (p_2)$ increase the number of small halo which can be a radiation source. In the left panel, the absorption occurs at $z> 30$, and hence the top left region cannot be constrained by EDGES result. In the right panel, reionization is completed earlier than $z=50$ around the top left corner and in this case  the position of absorption peak cannot be defined. Therefore the region on the top left corner in the right panel is shaded with gray. For large $M_{\rm turn}$, more halos cannot contribute to heating and ionization, and the absorption trough is peaked at lower redshift even for large values of $p_1$ or $p_2$. In such a case, the EDGES low-band result cannot constrain $p_1$ and $p_2$, however, such large $M_{\rm turn}$ region can be excluded by EDGES high-band data in which the redshift range sensitive to the observation is $z<12$.

Next we discuss how $L_X$, X-ray luminosity per star formation rate,
affects the constraints. 
In Fig.~\ref{fig:p1_Lx}, the constraint on the $p_1$--$L_X$ and $p_2$--$L_X$ planes are respectively shown in the left and right panels. In the figure, the value of $p_2$ ($p_1$) is fixed as $p_2=1$ ($p_1=1$)  in the left (right) panel. When $L_X$ is assumed to be the fiducial value of $L_X = 10^{40.5}~{\rm erg}~ {\rm s}^{-1} M_{\rm sun}^{-1}~  {\rm yr}$  as in Fig.~\ref{fig:result1}, the model with $(p_1,p_2) = (15,1.0)$ is ruled out because the 21cm GS is lower than $-75$~mK at $z>22$, which comes from the fact that the spin temperature is coupled early with gas temperature due to the WF effect and the gas is colder than the CMB temperature. However, this model can be allowed if  $L_X$ is larger than the fiducial value such as $L_X \simeq 10^{42}~{\rm erg}~ {\rm s}^{-1} M_{\rm sun}^{-1} ~{\rm yr}$. With such a large value of $L_X$,  the X-ray can heat the gas earlier than $z=22$, and hence the absorption trough gets shallower than the threshold value of $-75$~mK and cannot be constrained. However, we note that $L_X$ can be constrained by the combination of future 21cm fluctuations and luminosity function, from which the degeneracy is expected to be broken.

\begin{figure*}
\resizebox{160mm}{!}{
\includegraphics[width=10cm,bb= 0 0 648 576]{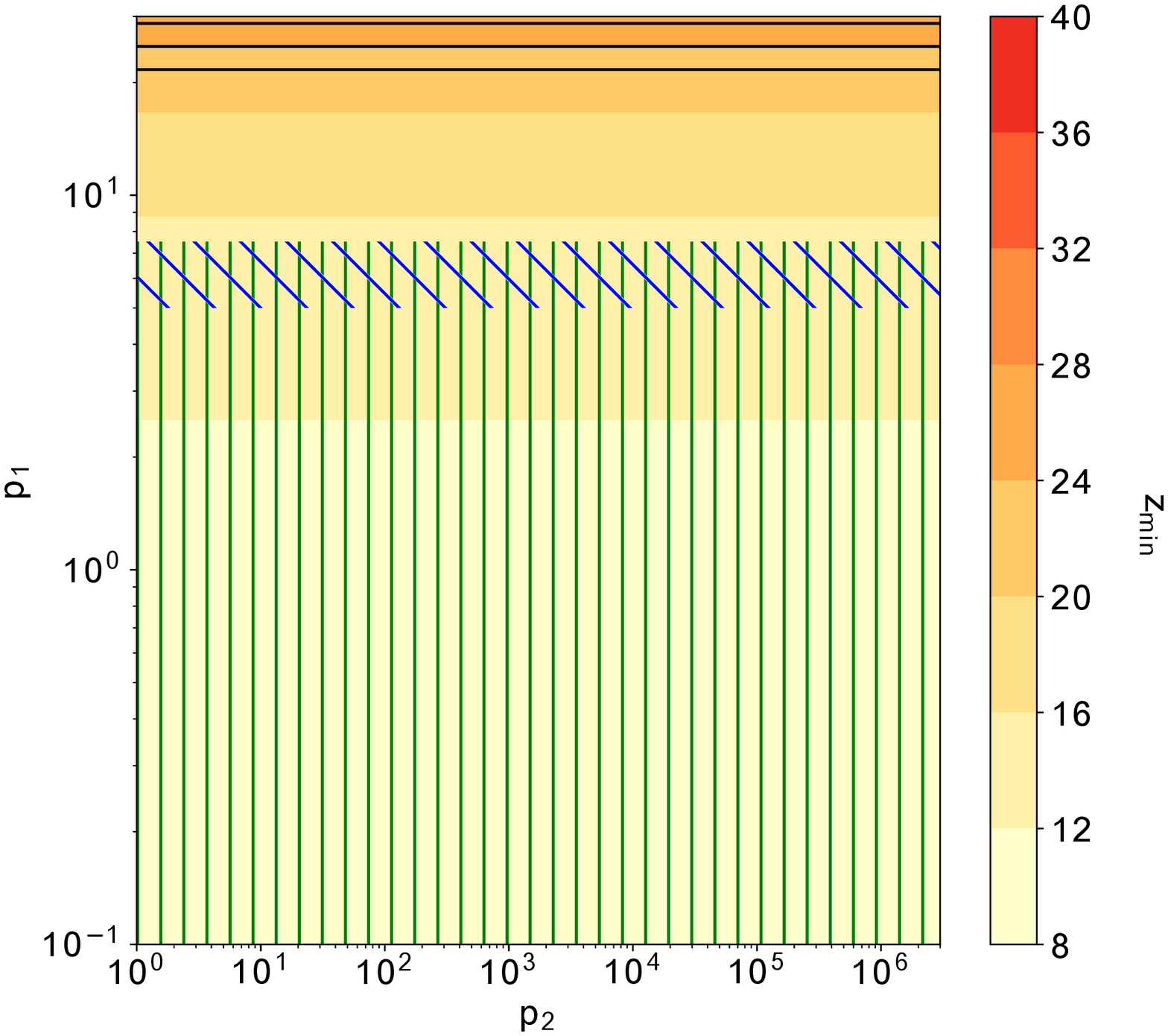} 
\includegraphics[width=10cm,bb= 0 0 648 576]{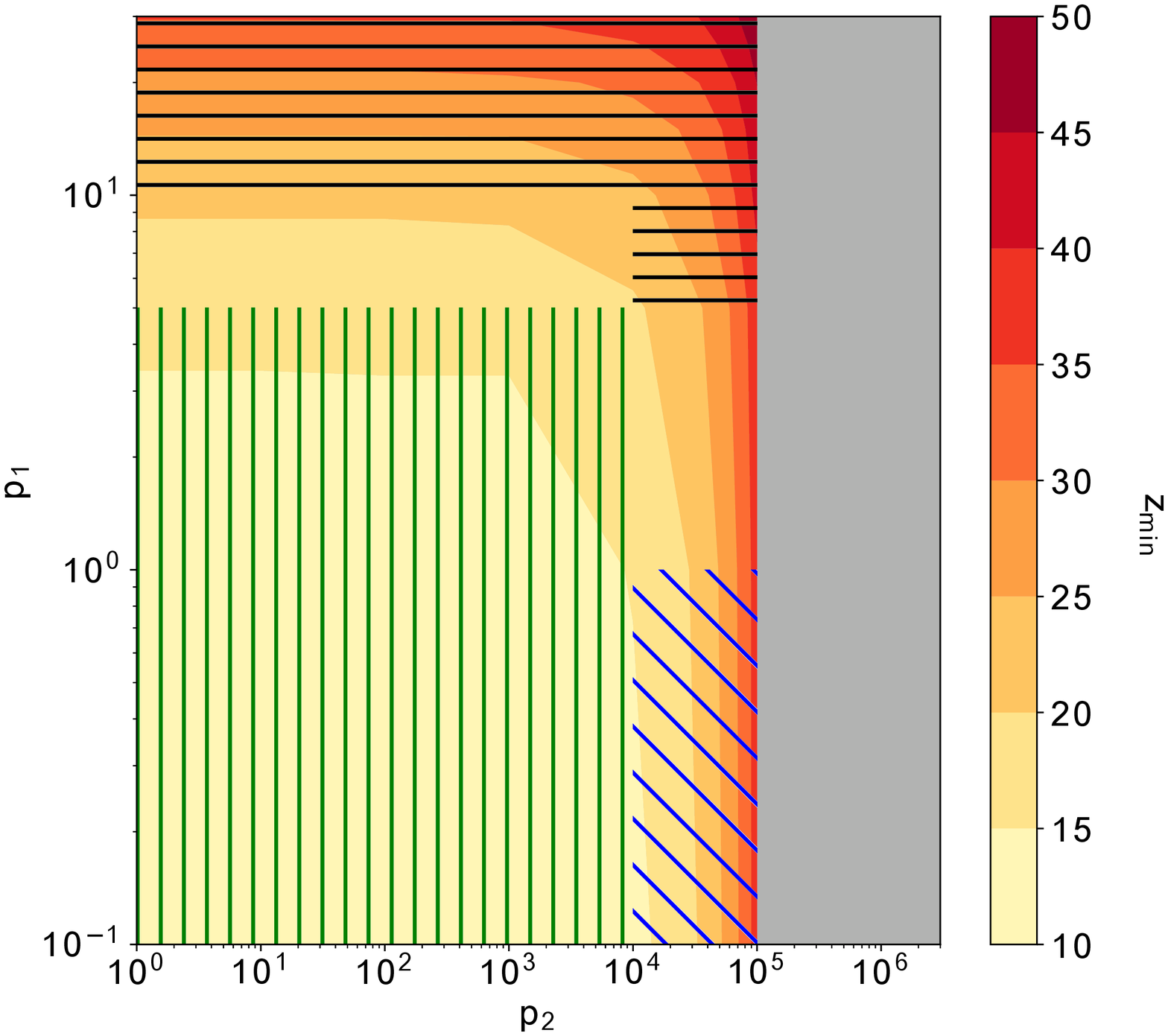}
}
\caption{\label{fig:result3} Constraints in the $p_1$-$p_2$ plane with $M_{\rm turn}=5\times 10^8 ~M_{\rm sun}$ where we adopt the Gaussian filter (left) and the real space top-hat filter (right) in calculating the HMF. The region filled by diagonal, horizontal and vertical lines are the same as Fig.~\ref{fig:result1}. We cannot define $z_{\rm min}$ at $p_2\gtrsim 10^5$ in the right panel because the ionization and heating happen before $z>50$. Such  region is shaded with gray.}
\end{figure*}

\bigskip
Now we discuss the issue of how the filter function affects the constraint. As shown in Fig.~\ref{fig:dndm}, the filter is crucial to estimate the HMF. Therefore the choice of filter may change the 21cm GS and the constraint on the PPS. In Fig.~\ref{fig:result3}, we show the results using different filters,  the Gaussian and the top-hat ones.  In left panel of Fig.~\ref{fig:result3}, we show the constraints on $p_1$ and $p_2$ using Gaussian filter in calculating the HMF. The Gaussian filter significantly reduces the power on small scales, and hence the $p_2$ cannot affect  the 21cm  GS. However, on the other hand, since the $p_2$ does not enhance the HMF, the degeneracy between $p_1$ and $p_2$ does not exist, and the 21cm GS can constrain the amplitude of PPS as $8\lesssim p_1 \lesssim 20$ from EDGES low-band data. In the right panel of Fig.~\ref{fig:result3}, we show the constraints on $p_1$ and $p_2$ using top-hat filter. The response to $p_1$ is similar to that for the smooth-$k$ filter. However, for the model with $p_2>10^6$, the ionization is completed before $z=50$ and the spin temperature is saturated. This is because large values of $p_2$  can cause an artificial feature of the HMF as shown in Fig.~\ref{fig:dndm}. 

It should be noted that the filter predicting the largest HMF in most scales is the top-hat one, followed  by those with the smooth-$k$  and the gaussian ones for the same PPS. Thus, the constraints with the top-hat filter is most stringent among three filters. We also mention that the difference in HMFs with these filters gets bigger for large values of $p(k)$. Thus, the difference in the constraints becomes large for high $p(k)$ models. However, it should be emphasized that, although, as seen from Fig.~\ref{fig:result3}, the constraints on $p(k)$ is affected by the choice of the filter function, the constraint on $p_1$ does not change much regardless of the filter: we obtain the constraint $p_1 \lesssim 10$ for any filter function. 
On the other hand, the constraint on $p_2$ can be much more affected. However, the limit on $p_2$ is generally weak for every filter.

\subsection{Constraints on Parametrization II}

\begin{figure*}
\resizebox{160mm}{!}{
\includegraphics[width=10cm,bb= 0 0 648 576]{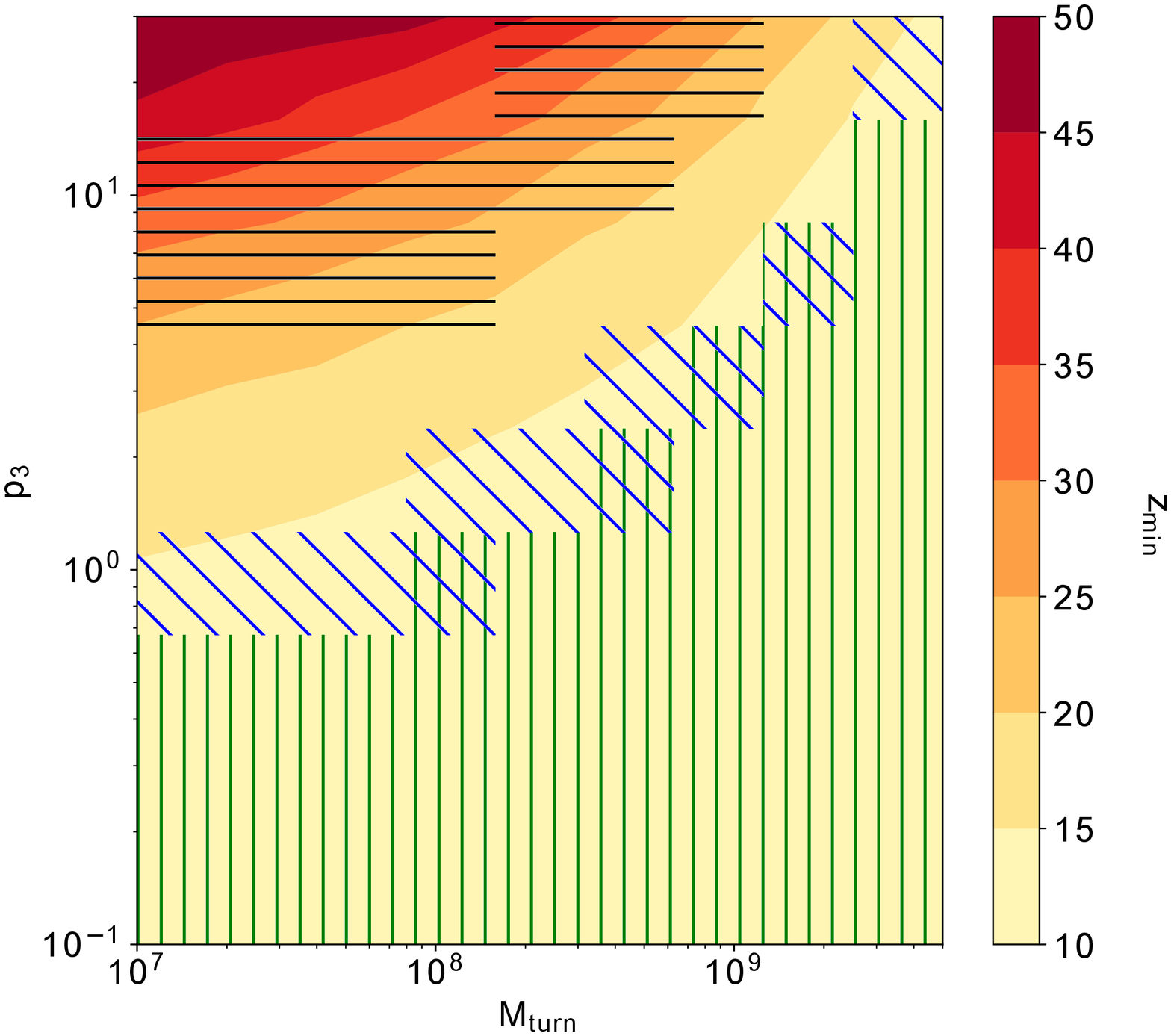} 
\includegraphics[width=10cm,bb= 0 0 648 576]{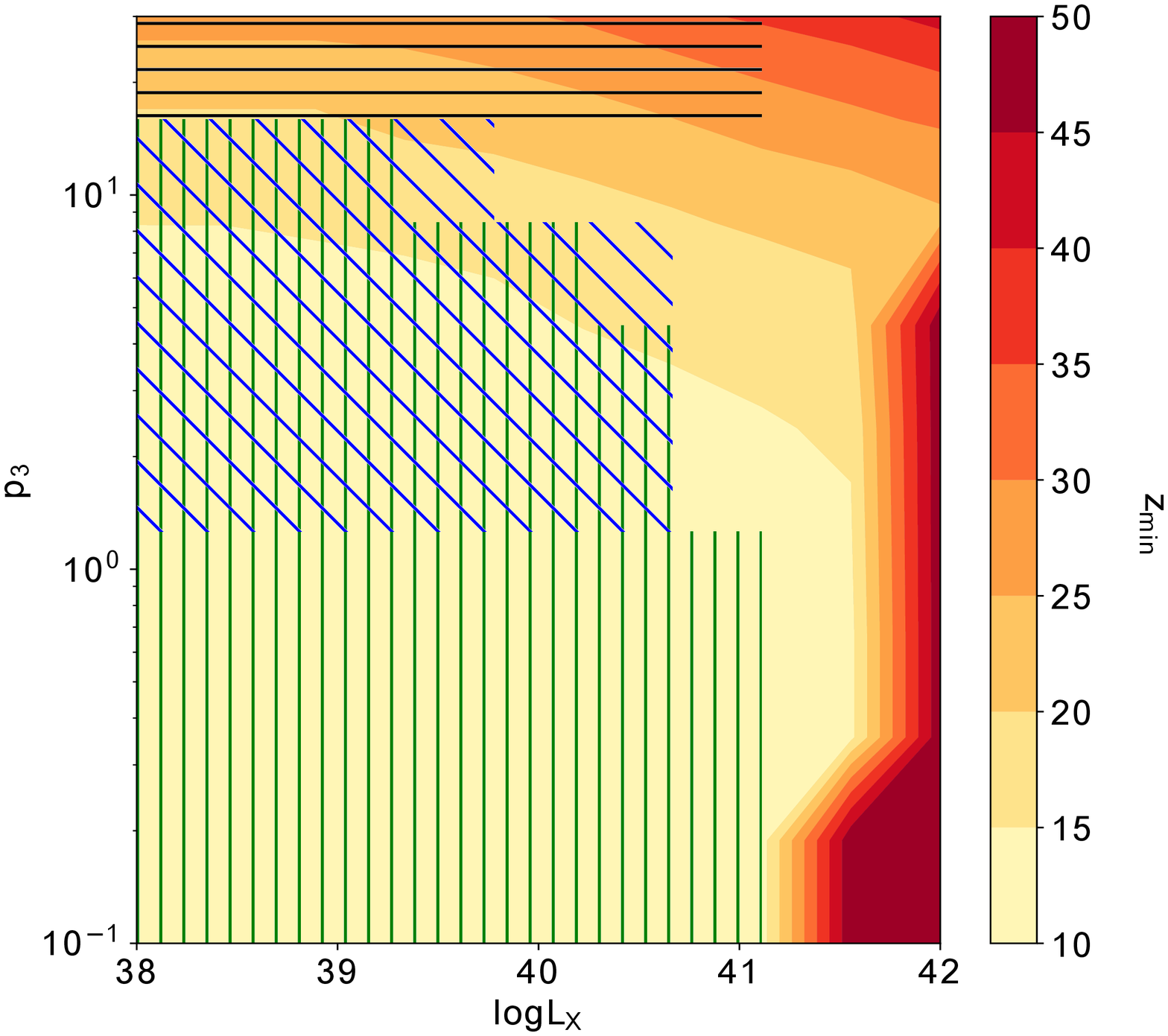}
}
\caption{\label{fig:p3} Constraint from EDGES on the $p_3$--$M_{\rm turn}$ (left) and $p_3$--$L_X$ (right) planes.  
}
\end{figure*}

Now in this section, we show constraints on  $p_3$ for the parametrization~II which corresponds to  the amplitudes at $10~{\rm Mpc}^{-1}<k<10^3~{\rm Mpc}^{-1}$.  
In Fig.~\ref{fig:p3}, constraints from EDGES on the $p_3$--$M_{\rm turn}$ (left panel) and $p_3$--$L_X$ (right panel) are shown in the same manner as Figs.~\ref{fig:result1}--\ref{fig:result3}.  The values of $L_X$ and $M_{\rm turn}$ are fixed as 
$L_X = 10^{40.5}~{\rm erg}~ {\rm s}^{-1} M_{\rm sun}^{-1}~  {\rm yr}$ and $M_{\rm turn} =5 \times 10^8 M_{\rm sun}$ in the left and right panels, respectively. Just as the left panel in Fig.~\ref{fig:result2}, the top left region in the left panel of Fig.~\ref{fig:p3}  cannot be constrained by EDGES since the absorption trough appears at $z>30$, which is outside the EDGES observation redshift.

According to the analysis for the parametrization~I, the amplitude of PPS at the scale of $10~{\rm Mpc}^{-1} < k <10^2~{\rm Mpc}^{-1} $, represented by $p_1$, is much more important for the 21cm GS than that at $10^2~{\rm Mpc}^{-1} < k <10^3~{\rm Mpc}^{-1} $, represented by $p_2$. Therefore the constraint on $p_3$ would be mostly determined by the amplitude on $10~{\rm Mpc}^{-1} < k <10^2~{\rm Mpc}^{-1}$ and it is severely constrained as $2  \lesssim p_3 \lesssim 8$ when $M_{\rm turn}$ and $L_X$ are fixed to be their fiducial values. However, as discussed in the previous section, there is a strong degeneracy between the small scale amplitude of PPS and astrophysical parameters such as $M_{\rm turn}$ and $L_X$ and the constraint on $p_3$ depends on the assumption on these parameters, which can  be seen from Fig.~\ref{fig:p3}. In particular, when one assumes a large values for $L_X$ as $L_X > 10^{42}~{\rm erg}~ {\rm s}^{-1} M_{\rm sun}^{-1} ~{\rm yr} $, we do not obtain any constraint on $p_3$. Nevertheless, as already mentioned,  it should be noted  that the combinations of observations of 21cm power spectrum and luminosity function can constrain $L_X$ severely, which would remove the degeneracy and in turn derive a more rigorous bound on the small scale amplitude of PPS.

\begin{figure*}
\resizebox{160mm}{!}{
\includegraphics[width=10cm,bb= 0 0 864 504]{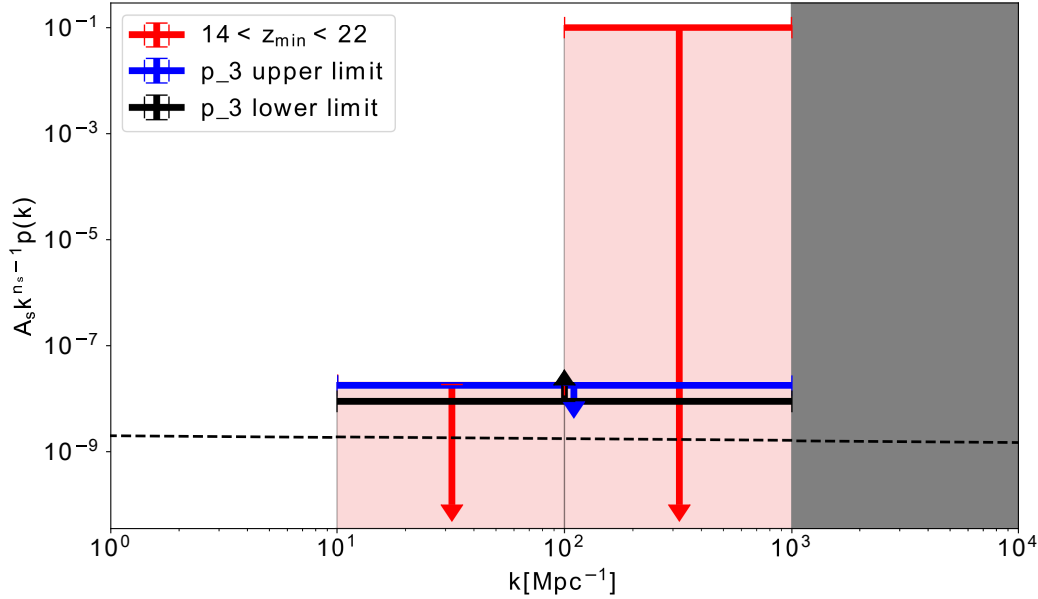}
}
\caption{\label{fig:result4} 
Bounds on the amplitude of primordial power spectrum on small scales. {To show the power of 21cm global signal to constrain the PPS, we adopt the EDGES results for illustration purposes.}
Black dashed line corresponds to the standard PPS extrapolated from large scales with $n_s=0.97$. Red arrows indicate the constraint on $p_1$ and $p_2$ obtained from EDGES low-band results. EDGES high-band (low-band) can put lower (upper) bounds on $p_3$ which are shown as Black (Blue) allow.  Notice that at larger scales where $k<10~{\rm Mpc}^{-1}$, the PPS is already constrained by current observations such as CMB, large scale structure and Lyman-$\alpha$ (see, e.g. Fig.~6 in \cite{Bringmann:2011ut}). Gray region is not considered in this paper. 
{To obtain the bounds, we have fixed astrophysical parameters as $M_{\rm turn}=5\times 10^8~ M_{\rm sun}$ and  $L_X = 10^{40.5}~{\rm erg}~ {\rm s}^{-1} M_{\rm sun}^{-1}~{\rm yr}$ and other  parameters are fixed to be the default values adopted in 21cmFAST. We emphasize that the constraint depends on astrophysical parameters. We should also bear in mind that there are other uncertainties such as redshift evolution of astrophysical parameters, alternative stellar population and additional physics which can explain too deep absorption trough reported by EDGES. This strong constraints on the PPS rely on the good knowledge on astrophysical parameters, which can be achieved by combining the 21cm power spectrum and 21cm GS once 21cm power spectrum is measured at $z>$14.  }
}
\end{figure*}

\bigskip
\bigskip

Finally we summarize the bounds on the small scale amplitude we obtained. In Fig.~\ref{fig:result4}, we show the lower and upper bounds on the amplitude of PPS at small scales for the case with the fiducial values of $M_{\rm turn}=5\times 10^8~ M_{\rm sun}$ and  $L_X = 10^{40.5}~{\rm erg}~ {\rm s}^{-1} M_{\rm sun}^{-1}~{\rm yr}$
adopting the smooth-$k$ filter. On the scale of $ 10 ~{\rm Mpc}^{-1}< k< 10^2~{\rm Mpc}^{-1}$, the upper limit is obtained as $p_1\lesssim10$ from EDGES low-band results.  This constraint is much tighter than other probes of small scale amplitude of PPS. Although the assumption on the astrophysical parameters can change the result to some extent, this shows a power of the 21cm GS in deriving a severe constraint on PPS. On the other hand, the constraint on $p_2$ is not so strong since the PPS at the scale of $ 10^2 ~{\rm Mpc}^{-1}< k< 10^3~{\rm Mpc}^{-1}$ cannot affect enough the number of halo which contribute to Lyman-$\alpha$ radiation.  On the other hand, at $k<10 ~\rm Mpc^{-1}$, the PPS has already been tightly constrained by the CMB, large scale structure and Lyman-$\alpha$ observations. 

It should be emphasized that the 21cm GS is also able to provide a lower limit on the small scale amplitude of PPS. Not only the EDGES low-band result but also high-band result would be useful to provide a  lower bound on the amplitude of PPS.  Although we could not obtain a lower bound for the case of Parametrization~I since the suppression of $p_1$ can be compensated by the enhancement of $p_2$ and vice versa. However, 
in the analysis for  Parametrization~II, we can obtain the lower bound on  the amplitude for $10~{\rm Mpc}^{-1} < k < 10^3~{\rm Mpc}^{-1}$ as $p_3 \gtrsim 2$ for the case where the fiducial values are assumed for $M_{\rm turn}$ and $L_X$. As already mentioned,  the constraint is very degenerate with the astrophysical parameters. Nevertheless those astrophysical parameters can also be constrained by other observations such as UV luminosity function, fluctuations of 21cm and so on. Therefore the degeneracy can be removed in future, which allows us to put a severer  constraint on the small scale amplitude of PPS.

When we adopt the parametrization~I, we have an additional degeneracy between $p_1$ and $p_2$. However, this degeneracy can be broken in tandem with other constraints. For example, the arguments of primordial black holes can constrain the small scala amplitude of PPS around $10~{\rm Mpc}^{-1} < k < 10^3~{\rm Mpc}^{-1}$ as $P_{\rm prim}  \lesssim 10^{-2}$ \cite{Josan:2009qn}, which motivates us to assume a prior of $p_1, p_2 < 10^7$. In particular, this prior on $p_2$ gives $5\lesssim p_1 \lesssim 10$ for $M_{\rm turn} = 5 \times 10^8~M_{\rm sun}$ as shown in the right panel of Fig.~\ref{fig:result1}.

{Note that we fixed astrophysical parameters to obtain the result presented in Fig.\ref{fig:result4}. Astrophysics parameters can be well constrained by the 21cm line power spectrum as shown in \cite{Park:2018ljd}. In practice, the constraint on the PPS can be provided by combining the 21cm power spectrum and the 21cm GS. We mention that, however, there are additional uncertainties regarding the high-$z$ sources. For example, in this work, we use the 21cmFAST to model the 21cm GS, and assume that astrophysical parameters are constant in time and there is only single population of stars. However, we note that, for example, as shown in \cite{2019MNRAS.483.1980M}, the evolution of astrophysical parameters is required to realize the 21cm line absorption signal observed by the EDGES low-band result. If we introduce such an evolution, it should give more uncertainty in the constraints. {However, we have studied the degeneracy between PPS and astrophysical parameters within wide parameter space and checked that the uncertainties in astrophysics parameters do not change our constraints by order of magnitude.}  Moreover, once the 21cm power spectrum is measured at $z>$14, the parameters would be well constrained even if these parameters evolve with redshift.}

{The assumption of the stellar population would also give large uncertainties on the constraint on the PPS. In \cite{2019arXiv191010171M}, they have shown that popIII stars might explain the EDGES low-band result if X-ray emissivity of popIII stars is small enough to keep the IGM being cooled and black holes converted from popIII stars radiate radio emission which makes strong radio background at very high redshift. Since we focus on relatively small scale halo abundance, the uncertainties due to the popIII stars would be non-trivial. If popIII stars increase the number of Lyman-$\alpha$  and X-ray photons, the absorption line shift to high redshift. In fact, this effect is same as decreasing the $M_{\rm turn}$ in practice, and in such a case, the upper limit becomes severer as shown in Fig.~\ref{fig:result2}. If X-ray emission of popIII stars is strong and make the absorption trough shallower, the 21cm GS experiment cannot constrain the parameters from the EDGES non-detection result. Such an uncertainty is same as the effect of $L_X$ shown in Fig.~\ref{fig:p1_Lx}.} {Therefore the assumption of the stellar population can be translated to the effect due to the uncertainty in $L_X$ which has been taken into account in our analysis.}

{Although to obtain the constraint on the PPS, we only use the fact that the EDGES has not detected any absorption signal except the redshift range of $14 < z < 22$, the reported absorption trough has shown a large amplitude and indicates interesting possibilities. For example, high-$z$ radio background can explain the strong absorption. The amplitude of the 21cm GS mainly depends on the term of $1-({T_{\rm CMB}+T_{\rm RB}})/{T_{\rm S}}$, where $T_{\rm RB}$ is the brightness  temperature of radio background. If the radio background is generated by any high-$z$ object such as black holes as suggested in \cite{2019arXiv191010171M}, a high redshift end of the position of the absorption trough does not change because the radio background should be generated after the Lyman-$\alpha$ emission from first stars. However, this is not correct if the radio background is not stellar origin and in such a case, a low redshift end of the absorption shifts to lower redshift. The lower bound on the PPS can be put by the argument that the absorption line should not be shifted to too lower redshift, which means that the existence of additional radio back ground makes the lower bound on the PPS stronger. We also note that the radio background is one of ideas to explain the latest EDGES absorption. Other effects such as interaction between baryon and dark matter can shift the absorption line to lower redshift as shown in Fig.2 in \cite{Barkana:2018lgd}, and the parameters (e.g. dark matter particle mass and scattering cross sections) might give a degeneracy in the parameters used in this work. However, exploring such uncertainties due to new physics is out of scope of this paper. }

\section{Summary and Conclusion}

We have studied how we can probe small scale amplitude of primordial fluctuations by using the global signal of neutral hydrogen 21cm line. Since the global signal of 21cm line is affected by early structure formation, it should depend on primordial fluctuations, especially, those on small scales. 

We have argued that, just by looking at the position of the absorption trough of the 21cm global signal, we can constrain the amplitude of primordial power spectrum on small scales. We focused on the scales of $10~{\rm Mpc}^{-1} < k < 1000~{\rm Mpc}^{-1}$ and investigated the amplitude of primordial power spectrum on these scales by using EDGES low-band and high-band results.  We have considered two parametizations, given in Eqs.~\eqref{eq:param1} and \eqref{eq:param2} to obtain the constrains on the amplitudes. $p_1$ and $p_2$ for Parametrization~I are constrained as $ p_1 \lesssim 10$ and $ p_2 \lesssim 6\times 10^7$, which corresponds to $P_{\rm prim} (10 ~{\rm Mpc}^{-1} < k < 10^2~{\rm Mpc}^{-1})  \lesssim {\cal O}(10^{-8}) $ and $P_{\rm prim} (10^2 ~{\rm Mpc}^{-1} < k < 10^3~{\rm Mpc}^{-1})  \lesssim {\cal O}(10^{-1}) $ when the fiducial values are assumed for $M_{\rm turn}$ and $L_X$.  Notice that the constraint on the amplitude around the scales of $10 ~{\rm Mpc}^{-1} < k < 10^2~{\rm Mpc}^{-1}$ is much severer than previous limits obtained from other methods. When we adopt Parametrization~II, one can obtain the bound on $p_3$ as $ 2 < p_3 <8$ which indicates that $P_{\rm prim}(10 ~{\rm Mpc}^{-1} < k < 10^{3}~{\rm Mpc}^{-1}) \sim {\cal O}(10^{-9}) - {\cal O}(10^{-8})$ for the fiducial values of $M_{\rm turn}$ and $L_X$.

However, we should note that the constraints strongly depend on astrophysical parameters assumed in the analysis as we mainly discussed this issue in Section~\ref{sec:param1}. In particular, for the case of large $L_X$, the 21cm absorption line becomes shallow, and the  EDGES results  cannot constrain the amplitude of PPS. Nevertheless despite some degenerate effects on the 21cm global signal among the amplitude of primordial fluctuations and astrophysical parameters, they can be removed by combining other observations.   Actually a recent analysis \cite{Park:2018ljd} has shown that the combination of 21cm power spectrum and UV luminosity function can severely constrain some of the astrophysical parameters, which can reduce the degeneracy.

It should also be mentioned that the constraints on the small scale amplitude  depends on the filter function too. We have adopted, as a benchmark,  the so-called smooth-$k$ filter function \cite{Leo:2018odn}, which is considered to give a good estimate especially for models with suppressed power spectrum on small scales. We have also studied other filter functions such as the top-hat and the Gaussian ones which were discussed in Section~\ref{sec:param1}.  Although we think that a smooth-$k$ filter function seems to to appropriate, we need to check the validity by performing $N$-body simulation, which is beyond the scope of this paper.  Although we have shown that the constraint also depends on the filter function, it should be emphasized that even if we adopt different filters, our final conclusion remains almost unchanged: $p_1$ is relatively well constrained as $p_1 < {\cal O}(10)$ and $p_2$ cannot be severely constrained. 

Given that the probes of small scale primordial fluctuations are limited, the global signal of 21cm line would be one of the important powerful tool to probe primordial fluctuation on small scales. Although, as mentioned above, the 21cm signal suffers from some uncertainties such as in astrophysical parameters in constraining the small scale amplitude of primordial power spectrum, it can constrain its amplitude just by looking at the position of the absorption trough. We can also utilize the whole shape of the 21cm global signal to probe primordial fluctuations, which is expected to give more information and more stringent constraints. This issue should also be worth investigating, which is left for a future work.

\section*{Acknowledgments}

T.T. is grateful to Yuichiro Tada for enlightening discussion. This work is partially supported by JSPS KAKENHI Grant Number 16H05999 (KT), 16J01585 (SY), 17H01110 (KT), 17H01131 (TT), 19K03874 (TT), MEXT KAKENHI Grant Number 15H05888 (TT), 19H05110 (TT), 15H05896 (KT), and Bilateral Joint Research Projects of JSPS (KT). SY is supported by JSPS Overseas 
Research Fellowships.

\bibliography{small_sclae_PS_21cm_global}

\end{document}